\documentclass[%
 reprint,
nofootinbib,
 amsmath,amssymb,
prd,
]{revtex4-1}

\usepackage{graphicx}
\usepackage{dcolumn}
\usepackage{bm}
\usepackage{hyperref}

\newcommand{\dif}{\mbox{d}}
\newcommand{\odelta}{\overline{\delta}}
\newcommand{\va}{\vec{v}\cdot\vec{a}}
\newtheorem{proposition}{Proposition}
\newtheorem{lemma}{Lemma}

\usepackage{color}

\begin{document}

\preprint{ICCUB 16-041, UTTG-25-16}

\title{Lie Symmetries of Non-Relativistic and Relativistic Motions}

\author{Carles Batlle}
\email{carles.batlle@upc.edu}
\affiliation{
Departament de Matem\`atiques and IOC, 
Universitat Polit\`ecnica de Catalunya\\
 EPSEVG, Av. V. Balaguer 1, E-08808 Vilanova i la Geltr\'u, Spain
}%

\author{Joaquim Gomis}%
\email{joaquim.gomis@ub.edu}
\affiliation{%
Departament de F\'{\i}sica Qu\`antica i Astrof\'{\i}sica and Institut de Ci\`encies del Cosmos\\
Universitat de Barcelona, Mart\'i i Franqu\`es 1, E-08028 Barcelona, Spain
}%

\author{Sourya Ray}%
\email{ray@uach.cl}
\affiliation{%
	Instituto de Ciencias F\'isicas y Matem\'aticas, Universidad Austral de Chile, Valdivia, Chile
}%

\author{Jorge Zanelli}%
\email{z@cecs.cl}
\affiliation{%
Centro de Estudios Cient\'ificos (CECs), Av. Arturo Prat 514, Valdivia, Chile
}%

\date{\today}

\begin{abstract}
We study the Lie symmetries of non-relativistic and relativistic higher order constant motions in $d$ spatial dimensions, \textit{i.e.} constant acceleration, constant  rate-of-change -of-acceleration (constant jerk), and so on. In the non-relativistic case, these symmetries contain the $z=\frac 2N$ Galilean conformal transformations, where $N$ is the order of the differential equation that defines the constant motion. The dimension of this group grows with $N$.

In the relativistic case the vanishing of the  ($d+1$)-dimensional space-time relativistic acceleration, jerk, snap, \ldots, is equivalent, in each case, to the vanishing of a $d$-dimensional spatial vector. These vectors are the   $d$-dimensional non-relativistic ones plus additional terms that guarantee the relativistic transformation properties of the corresponding $d+1$ dimensional vectors. In the case of acceleration there are no corrections, which implies that the Lie symmetries of zero acceleration motions are the same in the non-relativistic and relativistic cases. The number of Lie symmetries that are obtained in the relativistic case does not increase from the four-derivative order (zero relativistic snap) onwards. We also deduce a recurrence relation for the spatial vectors that in the relativistic case characterize the constant motions.
\begin{description}

\item[PACS numbers]  02.20Sv\  03.30.+p\   11.30.-j\  
 
\end{description}
\end{abstract}

\pacs{Valid PACS appear here}
\maketitle


\section{\label{sec:level1}Introduction}

The study of the symmetries of non-relativistic and relativistic motions in flat space time has been the subject of interest through the years, see \cite{Hill:1945}   \cite{Rindler:1960zz}. In particular, the motions with constant acceleration  in the relativistic case are at the basis of the Unruh effect \cite{Unruh:1976db}. The generalization of jerk and snap has attracted recent interest \cite{Dunajski:2008tg} \cite{Russo:2009yd} \cite{Pons:2018lnw}.

{The Lie   symmetries associated to a system of differential equations 
	\begin{equation} \label{q^(n)}
	L_a\left(t, q^A,\frac{dq^A}{dt}, \cdots, \frac{\dif ^n q^A}{\dif t^n}\right) {=0}, 
	\end{equation}
	for $a=1, \cdots, r$, $A=1,\cdots M$, 
	can be understood as those total,   or passive, transformations  
	\begin{equation} \label{CBA2}
	q^A \to \tilde{q}^A = q^A + \delta q^A(t,q),\ \ t \to \tilde{t} = t + \delta t(t,q),
	\end{equation}
	that map solutions of (\ref{q^(n)}) into other solutions of (\ref{q^(n)}) \cite{Olver1993b}. In order to determine these transformations it is useful to consider the associated functional,  or active, variation of $\overline{\delta} q^A$, defined as 
	\begin{equation}\label{CBA3}
	\overline{\delta} q^A = \delta q^A - \dot{q}^A \delta t,
	\end{equation}
	where the dot denotes the derivative with respect to $t$. Using the fact that the functional variation $\overline{\delta}$ commutes with the derivative with respect to $t$, the above symmetry requirement is equivalent to the demand that $\overline{\delta}q^A$ satisfies
	\begin{equation}\label{CBA4}
	\left. L_a[t,\overline{\delta}q]\right|_{L[t,q]=0} =0, \ a=1,\ldots, r.
	\end{equation}
	This results in a set of partial differential equations for $\delta q(t,q)$ and  $\delta t(t,q)$, which, in the more complex cases, can be solved with the help of algebraic computer packages \cite{Rocha:2011}.

	We only consider spacetime Lie symmetries, that is, we do not admit the presence of derivatives of the space coordinates in the right-hand sides of equation (\ref{CBA2}). As we will discuss later, this allows us to interpret the symmetries in terms of transformations between equivalent observers. From the mathematical point of view one could, however, remove this restriction and consider a wider class of transformations, that is, B\"acklund-Lie symmetries of the equations of motion depending up to derivatives of the space coordinates of one order less than the order of the equations.

	In the non-relativistic case it has been proposed \cite{Duval:2011mi}, \cite{Gomis:2011dw}, that the symmetry algebra of those motions  contains the family of Galilean Conformal groups \cite{Henkel:1997zz}\cite{Negro1997}. One of the motivations of this work is to understand if there exists a family of relativistic conformal algebras that are  a symmetry of the motion of zero jerk, zero snap and their generalizations \cite{Russo:2009yd}. In this context we will often meet the $N$-Galilean Conformal Algebra  or NGCA for short, \cite{Negro1997}\cite{Negro1997b}\cite{Henkel:1997zz}\cite{Duval:2011mi}\cite{Gomis:2011dw}. In $d+1$ space-time dimensions the NGCA has dimension
	\begin{equation}
	{  3}+d (N+1) + \frac{d(d-1)}{2},
	\end{equation}
	where the $3$ corresponds to time translations, dilatations and expansions    that form the $Sl(2,R)$ group,
	the last term counts the rotations in $d$-space, and the intermediate term $d(N+1)$ corresponds to space translations, ordinary Galilean boosts, and $N-1$ generations of higher order Galilean boosts, all in $d$-space. 
	
	Through this paper we will consider continuous transformations of a differential equation as space-time symmetries in $d+1$ dimensions connecting two observers which describe 
	a particle with constant position, constant velocity, constant acceleration, and so on,
	a point of view that was 
	also taken in \cite{Hill:1945}.
	
	For instance, in the non-relativistic case, two observers  with reference frames given by space-time coordinates $(t,\vec{x})$ and $(T,\vec{X})$ which observe a particle moving with constant speed  
	\begin{equation}
	\frac{\dif \vec{x}(t)}{\dif t} = \vec{v},\quad \frac{\dif \vec{X}(T)}{\dif T} = \vec{V},
	\end{equation}
	where $\vec{x}(t)$ and $\vec{X}(T)$ are the corresponding trajectories of the particle and $\vec{v}$, $\vec{V}$ the corresponding constant velocities, or, equivalently,
	\begin{equation}
	\frac{\dif^2 \vec{x}(t)}{\dif t^2} = \vec{0},\quad \frac{\dif^2 \vec{X}(T)}{\dif T^2} = \vec{0},
	\end{equation}
	are connected by a   transformation of the form (\ref{CBA2})
	\begin{equation} \label{CBA2b}
	\vec{x} \to \vec{X} = \vec{x} + \delta\vec{x}(t,\vec{x}),\ \ t \to T = t + \delta t(t,\vec{x}),
	\end{equation}
	which, according to the discussion following (\ref{CBA2}), yields a functional variation
	\begin{equation}\label{CBA3b}
	\overline{\delta} \vec{x} = \delta \vec{x} - \dot{\vec{x}} \delta t,
	\end{equation}  
	satisfying
	\begin{equation}
	\left. \frac{\dif^2 \overline{\delta}\vec{x}(t)}{\dif t^2}\right|_{\ddot{\vec{x}} = \vec{0}} = \vec{0}.
	\end{equation}
	This can be repeated for higher order constant motions, such as constant acceleration, constant jerk (time derivative of acceleration) and so on. In the non-relativistic case, each constant motion is characterized as the zero-derivative of the next order, but this is not so straightforward in the relativistic case. It still makes sense, however, to study the symmetries of zero motions in the relativistic case, regardless of whether they can be interpreted as constant motions of the previous order or not, provided the suitable Lorentz invariant generalizations of jerk, snap and so on are used. For the jerk this was done in \cite{Hill:1945}, and a generalization to snap and beyond was presented for the first time  in \cite{Russo:2009yd}. 
	
	Although a recurrence relation was presented in \cite{Russo:2009yd} for these higher order $d+1$ space-time vectors, it turns out that the study of the symmetries of the corresponsing zero motions is better done in terms   of a $d$-spatial vector containing the components of the $d+1$-vector, such that the vanishing of the latter is equivalent to the vanishing of the former. These vectors are the  non-relativistic acceleration, non-relativistic jerk, etc., plus additional terms that guarantee the relativistic transformation properties of the corresponding $d+1$-vectors. This leads to a novel recurrence formula for these spatial vectors appearing in the $(d+1)$-relativistic objects.}

The paper is organized as follows. Section \ref{sectionNR} discusses in detail the Lie   symmetries of non-relativistic motions describing constant motions starting with constant position and going up to constant jerk (zero snap), although we also present the general result for arbitrary higher-order constant motions. {   Section \ref{sectionR}  starts with a summary of the well-known  construction of higher order derivative $d+1$ vectors  of the world-line trajectory of a relativistic particle in an arbitrary reference frame \cite{Russo:2009yd} and presents new results about  the new structure of those vectors. As it was done in the non-relativistic case, the Lie symmetries of those motions are computed in \ref{secLSR}.  Finally, Section \ref{sectionC} summarizes our results. Appendix \ref{appA} proves a Lemma used in the main text and Appendix \ref{appB} discusses some relations in the instantaneous rest frame of a particle.

\section{Lie Symmetries of non-relativistic motions}
\label{sectionNR}
The Lie symmetries of non-relativistic motions with constant acceleration (zero jerk) has been studied for a long time. In 1945, Hill  \cite{Hill:1945}  studied  for the first time the one-dimensional example, both in the relativistic and non-relativistic cases. Here we will review the case of $d$ spatial dimensions and then extend the analysis to constant rate-of-acceleration (zero snap), and also to the dynamics higher order vanishing derivatives.

\subsection{Non-relativistic constant position}
We start with the simplest case of constant position, which is somehow special, in that it exhibits an infinite number of point symmetries, and also has a relation with the Carrollian limit of the relativistic case
\cite{Bergshoeff:2014jla}.

Consider the equation non-relativistic zero-velocity in $d$ space dimensions
\begin{align} \label{cba1}
\frac{\dif x^i}{\dif t}=0, \quad i=1, \ldots, d.
\end{align}
We write the infinitesimal point transformations as 
\begin{align}\label{Liesymmetries}
&\delta x^i=  \xi^i(t, x),\\
&\delta t=f(t, x).
\end{align}
The functional variation is then given by 
\begin{equation}
\odelta x^i(t)=\xi^i(t, x)-\dot x^i f(t, x).
\label{cba2}
\end{equation}
The Lie symmetries are given by
\begin{equation}
0= \left.\frac{\dif }{\dif t}\odelta x^i\right|_{(\ref{cba1})} =
\partial_t \xi^i,
\end{equation}
and no condition is obtained for $f(t,x)$.  The total variation of $x^i$ is given thus by an arbitrary space diffeomorphism $\xi^i=\xi^i(x)$, while $t$ can be transformed by an arbitrary, space-dependent diffeomorphism $\delta t= f(t,x)$. This makes physical sense: space points can be mapped to arbitrary space points and provided the map does not depend on time one still gets a fixed point, while time can be arbitrarily transformed at each point of space. This infinite set of symmetries appears in the study of the symmetries of a Carroll particle \cite{Bergshoeff:2014jla}}.

\subsection{Nonrelativistic constant velocity}
The equation of a non-relativistic particle with constant velocity in $D$ dimensions is
\begin{align} \label{d2}
\frac{\dif ^2 x^i}{\dif t^2}=0, \quad i=1, \ldots,  d.
\end{align}

We now consider the point  transformations (\ref{Liesymmetries}). The condition that they are Lie
symmetries is given by 

\begin{eqnarray}
0&=& 
\left.\frac{\dif^2 }{\dif t^2}\odelta x^i\right|_{(\ref{d2})}\nonumber \\
 &=& 
\partial_t^2 \xi^i + 2 \dot{x}^j \partial_t\partial_j\xi^i  
+ \dot{x}^k\dot{x}^j \partial_j\partial_k \xi^i\nonumber\\
& -& \dot{x}^i \partial_t^2 f  
- 2 \dot{x}^i\dot{x}^j \partial_t\partial_j f - \dot{x}^i\dot{x}^k\dot{x}^j \partial_j\partial_k f.
\end{eqnarray}		

This implies the set of independent equations
\begin{align}\label{x0}
&\partial_t ^2 \xi^i =0 \\
\label{x1}
& 2\partial_t \partial_j \xi^i = \delta^i_j \partial_t ^2 f \\
\label{x2}
& \partial_j \partial_k \xi^i = \delta^i_k \partial_t \partial_j f + \delta^i_j \partial_t \partial_k f  \\
\label{x3}
& \partial_j \partial_k f=0.
\end{align}
From (\ref{x0}) and (\ref{x1}) one finds
\begin{align}
\label{xi}
\xi^i &= a^i (x) + b^i (x) t\\
\label{f}
f &= \alpha(t) + \beta_i (t) x^i. 
\end{align}
Since equations (\ref{x1}) and (\ref{x2}) are second order in $x^i$ and $t$, the solutions are polynomials of second degree in these variables,
\begin{align}
\xi^i &= a_0^i + a^i{}_j x^j + \frac{1}{2}[\delta^i_j \beta_{1k} +\delta^i_k \beta_{1j}]x^k x^j + (b_0^i + \alpha_2 x^i)t \nonumber \\
f &= \alpha_0 + \alpha_1t + \alpha_2 t^2 +(\beta_{0i} + \beta_{1i} t)x^i. \label{symsNRZA}
\end{align}
where $a^i{}_j$ are the elements of a general $d\times d$ dimensional matrix.
The number of independent parameters is $(d+1)(d+3)$, which corresponds to the dimension of the projective group in $d+1$ dimensions, ${\cal P}_{d+1}$. The projective group consists of all transformations that map straight lines, which are the solutions to (\ref{d2}), to straight lines; these include rotations, translations and dilatations.  The Schr\"odinger group $Sch_{d+1}$
is a proper subgroup  of the projective group, and it is
the first element of the Lisfshitz Galilean Conformal algebras with dynamical 
exponent $ z=\frac{2}{N}$ \cite{Henkel:1997zz} \cite{Duval:2011mi}, in this case with $N=1$.

The vector fields which generate the total variations (\ref{symsNRZA}) are, for each parameter,
\begin{eqnarray}
\alpha_0 &:&\ \partial_t, \label{MRZA1}\\
\alpha_ 1 &:&\ t\partial_t, \label{MRZA2}\\
\alpha_ 2 &:&\  t(t\partial_t+ x^j\partial_j)\equiv A_2, \label{MRZA3}\\
a_0^i &:& \ \partial_i, \label{MRZA4}\\
b_0^i &:&\ t\partial_i, \label{MRZA5}\\
a^i_{\, j} &:& x^j\partial_i, \label{MRZA6}\\
\beta_{0i} &:&\ x^i \partial_t, \label{MRZA7}\\
\beta_{1i} &:& x^i (t\partial_t + x^j\partial_j) \equiv B_1^i, \label{MRZA8}
\end{eqnarray}
and they form the closed algebra given in Table \ref{table_NRA}.
Notice that by combining (\ref{MRZA5}) and (\ref{MRZA7}) one obtains the generators of Lorentz boost transformations. In fact, as we will see in Section \ref{secLSR}, the motions corresponding to zero relativistic acceleration are determined by exactly the same equation (\ref{d2}) of the non-relativistic case.
This result is related to the classification of kinematical algebras \cite{Bacry:1968zf} in the flat case.

\begin{table*}[t]
\onecolumngrid
	\begin{ruledtabular}
\begin{center} 
	\begin{tabular}{c| cccccccc}
		& $\partial_t$ & $t\partial_t$ & $A_2$ & $\partial_j$ & $t \partial_j$ & $x^k \partial_j$ & $x^j \partial_t$ & $B_1^j$  \\
		\hline 	
		$\partial_t$ & $0$ & $\partial_t$ & $2t\partial_t +x^j\partial_j $ & $0$ & $\partial_j$ & $0$ & $0$ &   $x^j \partial_t$              \\            
		$t\partial_t$ &  & 0 & $A_2$ &  0 & $t\partial_j$ & 0 & $-x^j \partial_t$ &   0\\ 
		$A_2$   & & & 0 & $-t\partial_j$ & 0  & 0 & $-B_1^j$ & 0 \\
		$\partial_i$ & & & & 0 & 0  & $\delta_i^k \partial_j$ & $\delta_i^j\partial_t$ & $\delta_i^j t\partial_t +\delta_i^j x^k\partial_k + x^j\partial_i $ \\ 
		$t \partial_i$ & & & & & 0  & $\delta_i^k t\partial_j$ & $\delta_i^j t \partial_t - x^j\partial_i$ & $\delta_i^j A_2$  \\ 
		$x^l \partial_i$ & & & & &  & $\delta_i^k x^l \partial_j - \delta^l_j x^k \partial_i$ & $\delta_i^j x^l \partial_t$ &  $\delta_i^j B_1^l$\\
		$x^i \partial_t$ &  & & & &  & & 0 & 0 \\
		$B_1^i$ & 	& & & &  & & &  0
	\end{tabular}
\end{center}
\end{ruledtabular}
\caption{Commutators of the Lie symmetry vector fields of  (\ref{d2}), with $A_2=t^2\partial_t+ tx^j\partial_j$.}
\label{table_NRA}
\end{table*}

\twocolumngrid
\subsection{Non-relativistic constant acceleration}

The condition for constant acceleration, that is, zero jerk, is
\begin{align} \label{d3}
\frac{\dif ^3 x^i}{\dif t^3}=0, \quad i=1, \cdots d
\end{align}

The symmetry conditions are now
\begin{align}
\label{NRca_eq}
\left.\frac{\dif ^3}{\dif t^3}\odelta x^i\right|_{(\ref{d3})} =0.
\end{align}

In contrast to the previous case, now $f$ cannot  depend on the spatial coordinates $x$. Indeed, since $\odelta x^i = \xi^i - \dot{x}^i f$, there is a unique term containing quadratic contributions from the accelerations, given by $-3\ddot{x}^i\ddot{x}^k \partial_k f$, when computing the left-hand side of (\ref{NRca_eq}). This implies
\begin{equation}
\partial_k f =0
\label{v20}
\end{equation}
and using this information the remaining terms yield the independent conditions
\begin{align}
& \partial_j\partial_k \xi^i = 0,\label{v21}\\
& \partial_t^3 \xi^i = 0, \label{v22}\\
& 3\partial_t^2 \partial_k \xi^i - \delta^i_{\ k} \partial_t^3 f = 0, \label{v23}\\
& \partial_t \partial_k \xi^i - \delta^i_{\ k} \partial_t^2 f = 0.\label{v24}
\end{align}

Combining these equations leads then to 
\begin{align}
f &= \alpha_0 +\alpha_1 t + \alpha_2 t^2,\nonumber\\
\xi^i &= a^i + a^i_j x^j + (b^i + 2 \alpha_2 x^i ) t + c^i t^2.
\label{symsd3}
\end{align}

{
The associated vector fields are
\begin{eqnarray}
\alpha_0 &:&\ \partial_t, \label{NRCA1}\\
\alpha_ 1 &:&\ t\partial_t, \label{NRCA2}\\
\alpha_ 2 &:&\  t(t\partial_t+ 2x^j\partial_j)\equiv A_3, \label{NRCA3}\\
a^i &:& \ \partial_i, \label{NRCA4}\\
b^i &:&\ t\partial_i, \label{NRCA5}\\
a^i_{\, j} &:& x^j\partial_i, \label{NRZC6}\\
c^i &:&\ t^2 \partial_i, \label{NRCA7}
\end{eqnarray}
and their commutators form the algebra given in Table \ref{table_NRCA}.    
It has $d^2+3d+3$ generators, and it  contains
the $N=2$ NGCA algebra without central extension\footnote{A dynamical realization of this algebra for $N=2$ with central extension was considered in \cite{Lukierski:2007nh} for $d=2$ and for general $d$ in \cite{Fedoruk:2011ua}, while the extension to arbitrary $N$ was presented in \cite{Gomis:2011dw}\cite{Galajinsky:2012rp}.}, with $d(d+1)/2$ extra generators corresponding to symmetric spatial linear transformations.
{  
	In contrast to the previous case of zero acceleration, now one cannot construct Lorentz boosts out of these generators, since there is no generator of the form $x^i\partial_t $.  
	The relativistic counterpart of (\ref{d3}), that is, the equation stating that the    
	$d+1$
	relativistic jerk is zero, has additional terms which restore Lorentz invariance of the corresponding $d+1$ vector, at the price of destroying most of the symmetries of (\ref{d3}).
}

\begin{table*}[t]
	\begin{ruledtabular}
	\begin{center} 
		\begin{tabular}{c| ccccccc}
			& $\partial_t$ & $t\partial_t$ & $A_3$ & $\partial_j$ & $t \partial_j$ & $x^k \partial_j$ & $t^2\partial_j$   \\
			\hline 	
			$\partial_t$ & $0$ & $\partial_t$ & $2t\partial_t +2x^j\partial_j $ & $0$ & $\partial_j$ & $0$ & $2t\partial_j$             \\            
			$t\partial_t$ &  & 0 & $A_3$ &  0 & $t\partial_j$ & 0 & $2t^2 \partial_j$ \\ 
			$A_3$   & & & 0 & $-2t\partial_j$ & $-t^2\partial_j$  & 0 & $0$ \\
			$\partial_i$ & & & & 0 & 0  & $\delta_i^k \partial_j$ & $0$  \\ 
			$t \partial_i$ & & & & & 0  & $\delta_i^k t\partial_j$ & $0$  \\ 
			$x^l \partial_i$ & & & & &  & $\delta_i^k x^l \partial_j - \delta^l_j x^k \partial_i$ & $-\delta_j^l t^2 \partial_i$ \\
			$t^2 \partial_i$ &  & & & &  & & 0
		\end{tabular}
	\end{center}
\end{ruledtabular}
	\caption{Commutators of the Lie symmetry vector fields of  (\ref{d3}), with $A_3=t^2\partial_t+ 2tx^j\partial_j$.}
	\label{table_NRCA}
\end{table*}
}

\subsection{Non-relativistic constant rate of change of acceleration and beyond}
The equations of motion for constant rate of acceleration, that is, constant jerk (or zero snap) are
\begin{align} \label{d4}
\frac{\dif ^4 x^i}{\dif t^4}=0, \quad i=1, \cdots, d.
\end{align}
The point Lie symmetry transformations satisfy
\begin{align}
\left.\frac{\dif ^4}{\dif t^4}\odelta x^i\right|_{(\ref{d4})} =0.
\end{align}
Applying the same procedure as in the previous cases one finds 
\begin{align}
f &= \alpha_0 +\alpha_1 t + \alpha_2 t^2 \\
\xi^i &= a^i + a^i_j x^j + (b^i + 3 \alpha_2 x^i ) t + c^i t^2 + d^i t^3,
\label{symsNRZS}
\end{align}
which contains $d^2+4d+3$ parameters.
{
The  vector fields associated to each parameter are $\partial_t$ ($\alpha_0$), $t\partial_t$ ($\alpha_1$), $A_4\equiv t^2 \partial_t + 3 t x^i\partial_i$ ($\alpha_2$), $\partial_i$ ($a^i$), $x^j\partial_i$ ($a^i_{\,j}$),  $t\partial_i$ ($b^i$), $t^2\partial_i$ ($c^i$), and $t^3\partial_i$ ($d^i$),
and they form the closed algebra given in Table \ref{table_NRCJ}.  This algebra contains the $N=3$ NGCA without central extension, but with $d(d+1)/2$ extra generators which, again, correspond to arbitrary symmetric spatial linear transformations.}

\begin{table*}[t]
	\begin{ruledtabular}
\begin{center} 
	\begin{tabular}{c| cccccccc}
		& $\partial_t$ & $t\partial_t$ & $A_4$ & $\partial_j$ & $t \partial_j$ & $x^k \partial_j$ & $t^2\partial_j$  & $t^3\partial_j$ \\
		\hline 	
		$\partial_t$ & $0$ & $\partial_t$ & $2t\partial_t +3x^j\partial_j $ & $0$ & $\partial_j$ & $0$ & $2t\partial_j$    &   $3t^2\partial_j$      \\            
		$t\partial_t$ &  & 0 & $A_3$ &  0 & $t\partial_j$ & 0 & $2t^2 \partial_j$  &  $3t^3\partial_j$\\ 
		$A_4$   & & & 0 & $-3t\partial_j$ & $-2t^2\partial_j$  & 0 & $-t^3\partial_j$ & $0$ \\
		$\partial_i$ & & & & 0 & 0  & $\delta_i^k \partial_j$ & $0$   & $0$\\ 
		$t \partial_i$ & & & & & 0  & $\delta_i^k t\partial_j$ & $0$  &  $0$\\ 
		$x^l \partial_i$ & & & & &  & $\delta_i^k x^l \partial_j - \delta^l_j x^k \partial_i$ & $-\delta_j^l t^2 \partial_i$ &  $-\delta_j^l t^3 \partial_i$  \\
		$t^2 \partial_i$ &  & & & &  & & 0  & 0 \\
		$t^3\partial_i $ & & & & & & &  & 0
	\end{tabular}
\end{center}
\end{ruledtabular}
\caption{Commutators of the Lie symmetry vector fields of  (\ref{d4}), with $A_4=\equiv t^2 \partial_t + 3 t x^i\partial_i$.}
\label{table_NRCJ}
\end{table*}

The symmetry transformations corresponding to higher order  zero dynamics, 
\begin{align} \label{dN}
\frac{\dif ^n x^i}{\dif t^n}=0, \quad i=1, \cdots d,\quad n\geq 5,
\end{align}
can also be computed, and one obtains
\begin{align}
f &= \alpha_0 +\alpha_1 t + \alpha_2 t^2\label{tN} \\
\xi^i &= a^i + a^i_j x^j + (b^i + (N-1) \alpha_2 x^i ) t +  \sum_{k=2}^{n-1} c_k^i t^{k}.\label{xN}
\end{align}
{   This contains the $N=n-1$ NGCA algebra without central extension, again with the extra $d(d+1)/2$ generators.}
This also encompasses the cases $n=3$ and $n=4$, but not $n=2$, which is a transition case from the infinite number of symmetries of the case $n=1$ and the regular case for $n\geq 3$. The number of symmetries is $d^2 + 4d +3$ for $n=2$ and $d^2+nd+3$ for $n\geq 3$. 

\section{A new look at higher order relativistic kinematics}
\label{sectionR}
Let us consider a $d+1$ Minkowski space-time. We denote by $\tau$ the proper time, $t$ is the lab time, and the corresponding derivatives are denoted by dots and primes, respectively. From the worldline $x^\mu(\tau)$ of a particle we construct
\begin{equation}
v^\mu = \dot{x}^\mu, \quad a^\mu = \dot{v}^\mu, \quad b^\mu = \dot{a}^\mu, \quad h^\mu = \dot{b}^\mu.
\end{equation}
In the lab  frame 
one has
\begin{equation}
x^\mu=(ct,\vec x).
\end{equation}
We define also
\begin{equation}
\vec{v} = \vec{x}\,', \ \ \vec{a} = \vec{v}\,', \ \ \vec{b} = \vec{a}\,', \ \ \vec{h} = \vec{b}\,'.
\end{equation}
Taking into account that
\begin{equation}
\frac{\dif}{\dif t} = \frac{1}{\gamma} \frac{\dif}{\dif\tau}, \quad \gamma=\left(1-\frac{\vec{v}\,^2}{c^2}\right)^{-1/2},
\end{equation}
one gets
\begin{eqnarray}
v^\mu &=& (\gamma c, \gamma \vec{v}),\\
a^\mu &=& (\gamma^4 \frac{\va}{c}, \gamma^4 \frac{\va}{c^2}\vec{v} + \gamma^2 \vec{a}),\\
b^\mu &=& (4\gamma^7 \frac{(\va)^2}{c^3}+ \frac{\gamma^5}{c}(\vec{a}\,^2 + \vec{v}\cdot\vec b),
\gamma^3\vec{b}+ 4\gamma^7 \frac{(\va)^2}{c^4}\vec{v} \nonumber\\
&+& \frac{\gamma^5}{c^2}(\vec{a}\,^2 + \vec{v}\cdot\vec b)\vec{v}+ 3 \gamma^5 \frac{\va}{c^2}\vec{a}),
\end{eqnarray}
where we have also used
\begin{equation}
\gamma' = \gamma^3 \frac{\va}{c^2}.
\end{equation}

From the expression of $v^\mu$ one has
\begin{equation}
v^\mu v_\mu=-c^2,    
\end{equation} 
and then, deriving with respect to $\tau$,
\begin{align}
v^{\mu}a_{\mu}=0,\\
a^{\mu}a_{\mu}+v^{\mu}b_{\mu}=0,\\
3a^{\mu}b_{\mu}+v^{\mu}h_{\mu}=0.
\end{align}
Note, in particular, that neither $b^\mu$ nor $h^\mu$ are orthogonal to the velocity. Following \cite{Russo:2009yd}, we define the relativistic jerk $j^\mu$ as 
the component of $b^\mu$ orthogonal to $v^\mu$,
\begin{equation}
j^\mu = b^\mu - \frac{b^\nu v_\nu}{v^2} v^\mu = b^\mu + \frac{b^\nu v_\nu}{c^2} v^\mu.
\label{rel_jerk}
\end{equation}
After some algebra one finds out that the temporal part of $j^\mu$ is given by
\begin{equation}
j^0 = 3 \gamma^7 \frac{(\va)^2}{c^3} + \gamma^5 \frac{\vec{v}\cdot\vec{b}}{c},
\end{equation}
while the spatial components are\footnote{$v^i$, $a^i$, $b^i$ denote the components of the Euclidean vectors $\vec v$, $\vec a$, $\vec b$, respectively, 
not the spatial components of $v^\mu$, $a^\mu$, $b^\mu$.}
\begin{equation}
j^i = \gamma^3 b^i + 3\gamma^5 \frac{\va}{c^2} a^i + 3 \gamma^7 \frac{(\va)^2}{c^4} v^i + \gamma^5 \frac{\vec{v}\cdot\vec{b}}{c^2}v^i.
\end{equation}
We define the relativistic snap, $s^\mu$, as the derivative  of $j^\mu$ (instead of $b^\mu$) with respect to $\tau$ and subtract again the component along $v^\mu$:
\begin{equation}
s^\mu = \frac{\dif j^\mu}{\dif\tau} - \frac{\frac{\dif j^\nu}{\dif\tau} v_\nu}{v^2} v^\mu =\frac{\dif j^\mu}{\dif\tau} + \frac{\frac{\dif j^\nu}{\dif\tau}v_\nu}{c^2} v^\mu .
\end{equation}
After some algebra this yields
\begin{equation}\label{rel_snap}
s^\mu = h^\mu - \frac{a^2}{c^2} a^\mu -3\frac{a^\nu b_\nu}{c^2}  v^\mu.
\end{equation}
The four vectors $a^\mu$, $j^\mu$, $s^\mu$, whose explicit expressions were given in \cite{Russo:2009yd}, are such that they have purely spatial components in the comoving frame, $v^\mu =0$, and
can be rewritten  in compact form as
\begin{eqnarray}
a^\mu &=& \left(\frac{\gamma^4}{c}\vec{v}\cdot\vec{A} , \gamma^2 \hat{M} \vec{A}\right),\label{cova}\\
j^\mu &=& \left(\frac{\gamma^5}{c}\vec{v}\cdot\vec{B}, \gamma^3 \hat{M} \vec{B}\right),\label{covb}\\
s^\mu &=& \left( \frac{\gamma^6}{c}\vec{v}\cdot\vec{H}, \gamma^4 \hat{M} \vec{H}\right).\label{covh}
\end{eqnarray}
Here we have defined
\begin{eqnarray}
\vec{A} &=& \vec{a}, \label{RA}\\
\vec{B} &=& \vec{b}+ 3\gamma^2 \frac{\va}{c^2}\vec a, \label{RB}\\
\vec{H} &=& \vec{h} + 6 \gamma^2 \frac{\va}{c^2} \vec{b} \nonumber\\
&+&
\left(
3\gamma^2 \frac{\vec{a}\,^2}{c^2} + 4\gamma^2 \frac{\vec{v}\cdot\vec{b}}{c^2} + 18 \gamma^4 \frac{(\va)^2}{c^4}
\right) \vec a  \label{RH}  \\
&=& \vec{B}\,' + 3\gamma^2 \frac{\va}{c^2}\vec{B} + \gamma^2 \frac{\vec{v}\cdot\vec{B}}{c^2}\vec{a}, \label{RH2}
\end{eqnarray}
and where the matrix $\hat M$ has components
\begin{equation}
\hat{M}_{ij}= \delta_{ij} + \frac{\gamma^2}{c^2} v_i v_j \, ,
\label{cbaA}
\end{equation}
whose inverse of is given by
\begin{equation} \label{cbaAi}
\hat{M}^{-1}_{ij} = \delta_{ij} - \frac{1}{c^2}v_iv_j \,.
\end{equation}

The appearance of the spatial vectors $\vec{A}$, $\vec{B}$, $\vec{C}$ in the components of the corresponding ($d+1$)-vectors $a^\mu$,  $j^\mu$,  $s^\mu$, 
together with the invertibility of $\hat M$, has interesting  consequences. In particular,  the vanishing of the  relativistic acceleration, jerk, snap,\ldots is equivalent, 
in each case, to the vanishing of the corresponding $d$-dimensional spatial vectors. 
These vectors are the $d$-dimensional non-relativistic ones plus additional terms that guarantee the relativistic transformation properties of the corresponding ($d+1$)-dimensional vectors. Explicitly,
\begin{eqnarray}
a^\mu =0 &\Leftrightarrow& \vec{A}=0,\\
j^\mu =0 &\Leftrightarrow& \vec{B}=0,\\
s^\mu =0 &\Leftrightarrow& \vec{H}=0.
\end{eqnarray}
From these expressions it also follows that
\begin{eqnarray}
\left.\vec{B}\right|_{\vec{A}=0} &=& 0,\\
\left.\vec{H}\right|_{\vec{B}=0} &=& 0.
\end{eqnarray}


As shown in \cite{Russo:2009yd}, the sequence of $(d+1)$-vectors $a^\mu$, $j^\mu$, $s^\mu$ can be extended to higher orders by means of the recurrence relation
\begin{equation}
P_{(n)}^\mu = \frac{\dif}{\dif\tau}P_{(n-1)}^\mu - \frac{P_{(n-1)}^\nu a_\nu}{c^2} v^\mu,
\label{rrPP}
\end{equation}
starting with $P_{(1)}^\mu=a^\mu$.  This recurrence relation preserves the property of orthogonality with $v$, since
\begin{equation}
v_\mu P_{(n)}^\mu = \frac{\dif}{\dif\tau} (v_\mu P_{(n-1)}^\mu ) =0
\end{equation}
provided $v_\mu P_{(n-1)}^\mu=0$. Using
\begin{equation}
v_\nu \frac{\dif}{\dif\tau} P_{(n-1)}^\nu = - a_\nu P_{(n-1)}^\nu,
\end{equation}
(\ref{rrPP}) can be rewritten as
\begin{equation}
P_{(n)}^\mu = \left(
\delta^\mu_{\ \nu} + \frac{v_\nu v^\mu}{c^2}
\right)
\frac{\dif}{\dif\tau} P_{(n-1)}^\mu.
\label{sr1}
\end{equation}

This allows for an interesting generalization of (\ref{cova}), (\ref{covb}), (\ref{covh}) that provides, in addition, a recurrence relation for the objects $\vec A$, $\vec B$, $\vec H$, \ldots

\begin{proposition}
\label{prop1}
The $(d+1)$-vectors $P_{(n)}^\mu$ defined by (\ref{sr1}), starting with $P_{(2)}^\mu=a^\mu$, can be written as
\begin{equation}
P_{(n)}^\mu = \left(
\frac{\gamma^{n+3}}{c} \vec{v}\cdot \vec{Q}_{(n)} , \gamma^{n+1} \hat M \vec{Q}_{(n)}
\right), \ n\geq 2,
\label{sr2}
\end{equation}	
with $\hat{M}$ the $d\times d$ matrix given in (\ref{cbaA}), and with the $d$-vectors $\vec{Q}_{(n)}$ satisfying the recurrence relation
\begin{equation}
\vec{Q}_{(n)} = \frac{1}{\gamma^n} \frac{\dif}{\dif t} \left( \gamma^n \vec{Q}_{(n-1)} \right) + \frac{\gamma^2}{c^2} (\vec{v}\cdot\vec{Q}_{(n-1)}) \vec{a},
\ n\geq 3,
\label{sr3}
\end{equation} 
starting with $\vec{Q}_{(2)} = \vec{A}$.
\end{proposition}

In order to prove this result we need the following lemma, which is proved in Appendix \ref{appA}.

\begin{lemma}
\label{lemma1}
Let
\begin{equation}
Y^\mu = \left(
\delta^\mu_{\ \nu} + \frac{v_\nu v^\mu}{c^2}
\right) X^\mu.
\label{sr4}
\end{equation}
Then $Y^\mu$ can be written as
\begin{equation}
Y^\mu = \left(
\frac{\gamma^2}{c} \vec{v}\cdot\vec{K}, \hat{M}\vec{K}
\right),
\label{sr5}
\end{equation}
with 
\begin{equation}
\vec{K} = \vec{Y} - \frac{Y^0}{c}\vec{v}.
\label{sr6}
\end{equation}
Furthermore, 
\begin{equation}
Y^2 = \vec{K}^2 + \frac{\gamma^2}{c^2} (\vec{v}\cdot\vec{K})^2\geq 0,
\label{sr7}
\end{equation}
or, equivalently,
\begin{equation}
v\cdot Y =0.
\label{sr71}
\end{equation}
\end{lemma}

\textbf{Proof of Proposition \ref{prop1} }
From (\ref{sr1}) it follows that $P_{(n)}^\mu$, obeys a relation of the type given in (\ref{sr4}), with $X^\mu=\frac{\dif}{\dif\tau}P_{(n-1)}^\mu$ for each $n$, and hence there exists a $\vec{K}_{(n)}$ such that it can be expressed as in  (\ref{sr5}), (\ref{sr6}). In particular
\begin{eqnarray}
K_{(n)}^i &=& P_{(n)}^i - \frac{v^i}{c}P_{(n)}^0 \nonumber\\
&=& \frac{\dif}{\dif\tau} P_{(n-1)}^i - \frac{\gamma}{c^2} P_{(n-1)}^\nu a_\nu v^i\nonumber\\  
&-& \frac{v^i}{c} \left(
\frac{\dif}{\dif\tau} P_{(n-1)}^0 - \frac{\gamma}{c} P_{(n-1)}^\nu a_\nu
\right)\nonumber\\
&=&  \frac{\dif}{\dif\tau}\left(
P_{(n-1)}^i -\frac{v^i}{c} P_{(n-1)}^0
\right) + \frac{1}{c} P_{(n-1)}^0 \frac{\dif v^i}{\dif\tau}\nonumber\\
&=& \frac{\dif}{\dif\tau} K_{(n-1)}^i + \frac{1}{c} P_{(n-1)}^0 \frac{\dif v^i}{\dif\tau}\nonumber\\
&=& \gamma \frac{\dif}{\dif t} K_{(n-1)}^i + \frac{\gamma}{c} \frac{\dif v^i}{\dif t} \frac{\gamma^2}{c} (\vec{v}\cdot \vec{K}_{(n-1)})\nonumber\\
&=& \gamma \frac{\dif}{\dif t} K_{(n-1)}^i  + \frac{\gamma^3}{c^2} (\vec{v}\cdot \vec{K}_{(n-1)}) a^i.
\label{sr81}
\end{eqnarray}
Introducing now $Q_{(n)}^i$ by means of
\begin{equation}
K_{(n)}^i = \gamma^{n+1} Q_{(n)}^i,
\label{sr82}
\end{equation}
equation (\ref{sr81}) for the $\vec{K}$s can be rewritten as
\begin{equation}
\gamma^{n+1} Q_{(n)}^i = \gamma \frac{\dif}{\dif t} \left(  \gamma^n Q_{(n-1)}^i  \right) + \frac{\gamma^{n+3}}{c^2} \left(   
\vec{v}\cdot \vec{Q}_{(n-1)}
\right) a^i
\end{equation}
which, after some algebra, becomes (\ref{sr3}). Finally,using (\ref{sr5}) for $P_{(n)}^\mu$ and expressing $\vec{K}_{(n)}$ in terms of $\vec{Q}_{(n)}$ yields (\ref{sr2}), and this concludes the proof.$\Box$
\vskip1mm

Using (\ref{sr2}) and (\ref{sr3}), the next term in the sequence $a^\mu$, $b^\mu$, $j^\mu$,  $s^\mu$ would be the relativistic "crackle" $k^\mu$, given by 
\begin{equation}
k^\mu = P^\mu_{(4)} = \left(
\frac{\gamma^7}{c} \vec{v}\cdot\vec{C}, \gamma^5 \hat{M} \vec{C}
\right),
\label{sr8}
\end{equation}
with the spatial vector $\vec{C}$ obtained from $\vec{H}$ by
\begin{eqnarray}
\vec{C}&=&\frac{1}{\gamma^4}\frac{\dif}{\dif t }\left(\gamma^4 \vec{H}  \right) + \frac{\gamma^2}{c^2}(\vec{v}\cdot\vec{H}) \vec{a}\nonumber\\
&=&  \frac{\dif\vec{H}}{\dif t }   + \frac{4}{\gamma} \frac{\dif\gamma}{\dif t }\vec{H} +\frac{\gamma^2}{c^2}(\vec{v}\cdot\vec{H}) \vec{a}.
\label{sr9}
\end{eqnarray}

As in the lower order cases, (\ref{sr3})  shows that, if $\vec{Q}_{(n)}$ vanishes, then $\vec{Q}_{(m)}$ is also identically zero for all $m>n$; furthermore, from (\ref{sr2}) the invertibility of $\hat{M}$ implies that $P_{(n)}=0$ if and only if $\vec{Q}_{(n)}=0$. On the other hand, one can also see from (\ref{sr3}) --or from (\ref{rrPP}) for that matter, that $\vec{Q}_{(n)}=\vec{0}$ does not necessarily imply that $\vec{Q}_{(n-1)}$ is constant, marking  a departure from the situation in the non-relativistic case.  In fact, the question of defining movements with constant jerk, snap and beyond is conceptually complex \cite{Russo:2009yd} \cite{Pons:2018lnw}. 
We will just present a short discussion, in the $1+1$ space-time case in Appendix \ref{appB}, using the $d$-dimensional spatial vectors we have encountered.

\section{Lie symmetries of higher relativistic dynamics}
\label{secLSR}
In this section we will compute the Lie symmetries of the equations of motion corresponding to zero relativistic acceleration, jerk, snap and beyond in an arbitrary lab frame. As commented in the Introduction, and following the ideas in \cite{Hill:1945}, these symmetries can be seen as the space-time transformations that connect different lab frames where the corresponding zero motion is preserved. 

\subsection{Symmetries of $\vec A=0$}

Since the relativistic zero acceleration is finally equivalent to $\vec{A}=0$, one has the equations
\begin{equation}
\label{eqA}
\frac{\dif^2 x^i}{\dif t^2}=0
\end{equation}
which are the same that those of the nonrelativistic zero acceleration case. This means that the set of Lie symmetries is given by the fields  (\ref{MRZA1} - \ref{MRZA8}). This is not the case for higher order zero motions, for which the relevant quantities exhibit extra terms (see (\ref{RB}) and (\ref{RH})).

\subsection{Symmetries of $\vec{B}=0$}
The equations of motion $B_i=0$ are 
\begin{equation}\label{eqB}
b_i + 3 \frac{v_k a_k}{c^2- v_k^2} a_i=0,\ \ i=1,\ldots, d,
\end{equation}

Proceeding as in the non-relativistic case, one finds out that the symmetries of (\ref{eqB}) are given by the $d(d-1)/2+3d+3$ vector fields
\begin{eqnarray}
\partial_t, \label{rj1}\\
\partial_i,\\
B_i\equiv M_{0i}= x_i\partial_t + c^2 t \partial_i,\\
M_{ij}=x_j\partial_i-x_i\partial_j, \\
D=t\partial_t + x^i \partial_i,\\
C_i=(c^2 t^2-x^jx_j)\partial_i + 2x_i (t\partial_t+x^j\partial_j),\\
C=(c^2 t^2+x^jx_j)\partial_t + 2c^2t x^j\partial_j,\label{rj7}
\end{eqnarray}
whose commutators, given in Table \ref{table_ZRB}, form the conformal algebra in $d+1$ Minkowski space-time. This result generalizes that of \cite{Hill:1945} in the one-dimensional case.

\begin{table*}[t]
	\begin{ruledtabular}
\begin{center} 
	\begin{tabular}{c| ccccccc}
		& $\partial_t$ & $\partial_j$ & $B_j$ & $M_{jk}$ & $D$ & $C_j$ & $C$  \\
		\hline 	
		$\partial_t$ & $0$ & $0$ & $c^2 \partial_j$ & $0$ & $\partial_t$ & $2B_j$ & $2 c^2 D$        \\            
		$\partial_i$ &  & 0 & $\delta_{ij}\partial_t$ &  $ \delta_{ij} \partial_k - \delta_{ik}\partial_j$ & $\partial_i$ & $2M_{ij}+2\delta_{ij}D $ & $2 B_i$  \\ 
		$B_i$   & & & $-c^2 M_{ij} $& $  \delta_{ij} B_k - \delta_{ik}B_j $ & $0$  & $\delta_{ij}C$ & $c^2 C_i$  \\
		$M_{il}$ & & & & $ -\delta_{lj} M_{ik}+ \delta_{lk}M_{ij}+ \delta_{ij}M_{lk}-\delta_{ik}M_{lj}    $ & 0  & $ \delta_{ij}C_l-\delta_{lj}C_i$ & $0$   \\ 
		$D$ & & & & & 0  & $C_j$ & $C$  \\ 
		$C_i$ & & & & &  & $0$ & $0$   \\
		$C$ &  & & & &  & & 0
	\end{tabular}
\end{center}
\end{ruledtabular}
\caption{Commutators of the Lie symmetry vector fields of  (\ref{eqB}).}
\label{table_ZRB}
\end{table*}

\subsection{Symmetries of $\vec H=0$}
Finally, the symmetries of the 4th order differential equations
\begin{equation}
\label{eqH}
H^i =0,\ \ i=1,\ldots, d,
\end{equation}
are described by the $d(d-1)/2+2d+2$ vector fields
\begin{eqnarray}
\partial_t,\\
\partial_i,\\
B_i=x_i\partial_t + c^2 t \partial_i,\\
M_{ij}=x_j\partial_i-x_i\partial_j,\\
D=t\partial_t+ x^i\partial_i,
\end{eqnarray}
whose commutators satisfy the Weyl algebra (the subalgebra  of the conformal algebra in $d+1$ given in the $5\times 5$  upper-left part of Table \ref{table_ZRB}).

\subsection{Symmetries of higher order zero motions}
The symmetries of higher order zero motions, 
\begin{equation}
P^\mu_{(n)}=0, \quad n\geq 5,
\end{equation}
which, according to  (\ref{sr2}), are those of 
\begin{equation}
\label{sr20}
Q^i_{(n)} =0,\ \ i=1,\ldots, d,\quad n\geq 5 \, ,
\end{equation}
already appearing for $n=4$, \textit{i.e.} in the the snap case. Thus, the group of symmetries remains the Weyl group for all higher order zero motions. This is in contrast with the non-relativistic case, were the algebra of symmetries grows due to the enlargement of the $NGCA$ algebra with $N$.

\section{Conclusions}

\label{sectionC}

Table \ref{tablesum} summarizes our results with respect to symmetry algebras of the zero dynamics that we have studied in this paper.
We have also considered $n=1$ in the non-relativistic case, which yields an infinite dimensional algebra and which is not shown in the table.

\begin{table*}[t]
	\begin{ruledtabular}
	\begin{center}
		\begin{tabular}{|c|l|l|}
			\hline
			\textsc{$n$-order zero dynamics} & \textsc{Non-relativistic} & \textsc{Relativistic} \\ \hline\hline
			$n=2$   & Projective in $d+1$ (contains $N=1$ $NGCA$) & Projective in $d+1$ \\ \hline
			$n=3$   & Extended$^*$ $NGCA$ $N=2$ in $d+1$  & Conformal in $d+1$  \\ \hline
			$n\geq 4$   &   Extended$^*$ $NGCA$ $N=n-1$ in $d+1$ &  Weyl in $d+1$\\ \hline
		\end{tabular}
	\end{center}
\end{ruledtabular}
	\caption{Algebra of Lie symmetries  corresponding to several non-relativistic and relativistic zero dynamics in $d+1$ space-time. The superscript $^*$ indicates $d(d+1)/2$ extra generators added to those of the standard $NGCA$.}
	\label{tablesum}
\end{table*}

Beyond $n=2$, the number of symmetries of zero motions in the non-relativistic case increases with $n$ 
because the number of generators in the $N$ Galilean Conformal Algebra  increases with $N$. In contrast to this, in the relativistic case the number of symmetries decreases with $n$, and this answers in the negative our initial question about the existence of a family of higher order relativistic conformal algebras. Conformal algebras only appear in the relativistic case for $n=2$ and $n=3$, the latter being the standard conformal algebra, and from this point onwards there is only a vestige of conformal symmetry in the form of space-time dilatations.

For $n=2$ the set of  Lie symmetries are the same for the relativistic and non-relativistc cases. For $n=1$ only the non-relativistic case can be considered, since the relativistic velocity $v^\mu=(\gamma c,\gamma \vec v)$ cannot be made equal to zero unless we take $c=0$. In fact, in the non-relativistic case, zero velocity motions, that is, constant position motions, have the same Lie symmetries as those of  the Carrollian limit ($c\to 0$) of a relativistic particle.

	Although we have limited ourselves to Lie symmetries of the zero motions, one can also study the Noether symmetries of the actions which, when available, yield the corresponding equations of motion. In general, the Noether symmetries form a subgroup of the Lie ones.  Such actions are easily available for non-relativistic even order zero motions, as is also the case for the zero acceleration relativistic movement; its existence and interpretation for odd order in the non-relativistic case and in the general relativistic one is under study.  Since the symmetries presented here are more
		general than those of the standard $NGCA$ it would also be interesting to use  the technique of nonlinear realizations (see, for instance, \cite{Alvarez:2007fw}\cite{Lukierski:2007nh}\cite{Fedoruk:2011ua}\cite{Galajinsky:2012rp}) to construct the corresponding actions.

	In the relativistic case one could also consider the Lie symmetries of the equations of motion when these are expressed using the proper time as the independent variable instead of the time in an arbitrary frame, adding the proper time condition as a new equation of motion. For zero acceleration the set of Lie symmetries boils down to the Weyl group, from the full projective group when using the arbitrary frame time. Such a reduction in the symmetry has also been reported, in the context of the equations of motion derived from a Lagrangian without fixing  the gauge, in
	\cite{Christodoulakis:2013xha}.

\section*{Acknowledgements}\addcontentsline{toc}{section}{Acknowledgements}
 
We would like to thank Avijit Lahiri, Anindya Ghose Choudhury, G. P. Sastry, Sayan Kar, Koushik Dutta, Maximilian Wentzel, Soumya Ray, Siddhartha Dechoudhury, Axel Kleinschmidt, Jorge Russo, Paul Townsend and Josep Maria Pons for useful discussions.

The work of CB has been partially supported by Generalitat de Catalunya through project 2017 SGR 872. 
JG has been supported in part by MINECO FPA2016-76005-C2-1-P and Consolider CPAN, and by the Spanish goverment (MINECO/FEDER) under project MDM-2014-0369 of ICCUB (Unidad de Excelencia Mar\'{i}a de Maeztu). JG has also been
supported by CONICYT under grant PAI801620047 as a visiting professor of the Universidad Austral de Chile.
SR is supported by FONDECYT grant 1150907.
JZ has been partially funded by FONDECYT grant 1180368. The Centro de Estudios Cient\'{i}ficos (CECs) is funded by the Chilean Government
through the Centers of Excellence Base Financing Program of Conicyt.


\begin{appendix}


\section{Proof of Lemma  \ref{lemma1} } 

\label{appA}

From (\ref{sr4}) one has
\begin{eqnarray}
Y^i &=& X^i + \gamma (X\cdot v)\frac{v^i}{c^2},\label{sr21}\\
Y^0 &=& X^0 + \gamma (X\cdot v)\frac{1}{c}\label{sr22}.  
\end{eqnarray}
Combining these one gets
\begin{equation}
Y^i - \frac{v^i}{c} Y^0= X^i-\frac{v^i}{c}X^0 \equiv K^i,
\label{sr23}
\end{equation}
where $K^i$ is a constant quantity, the same for all the elements of  the sequence of $d+1$ vectors given by (\ref{sr4}). From (\ref{sr22}) one now gets
\begin{eqnarray}
Y^0 &=& X^0-\gamma^2 X^0 + (\vec{X}\cdot\vec{v})\frac{\gamma^2}{c}\nonumber\\
&=&  (\vec{X}\cdot\vec{v})\frac{\gamma^2}{c} -\gamma^2 \frac{\vec{v}^2}{c^2}X^0\nonumber\\
&=& \frac{\gamma^2}{c}\vec{v}\cdot\left(       \vec{X}-\frac{X^0}{c}\vec{v}             \right)\nonumber\\
&=& \frac{\gamma^2}{c}\vec{v}\cdot\vec{K},
\label{sr24}
\end{eqnarray}
and then
\begin{eqnarray}
Y^i &=& K^i + \frac{v^i}{c}Y^0 \nonumber\\
&=& K^i + \frac{\gamma^2}{c^2} (\vec{v}\cdot\vec{K}) v^i.
\label{sr25}
\end{eqnarray}
Putting (\ref{sr23}) and (\ref{sr24})  together one has
\begin{eqnarray}
Y^\mu &=& \left(
\frac{\gamma^2}{c}\vec{v}\cdot\vec{K}, 
\vec{K} + \frac{\gamma^2}{c^2} (\vec{v}\cdot\vec{K}) \vec{v}.
\right)\nonumber\\
&=& \left(
\frac{\gamma^2}{c}\vec{v}\cdot\vec{K}, 
\hat{M}\vec{K}
\right),
\label{sr26}
\end{eqnarray}
with 
\begin{equation}
\hat{M}_{i j} = \delta_{i j} + \frac{\gamma^2}{c^2} v_i v_j,
\end{equation}
as desired. The last two results of the lemma follow then  from a simple calculation.

\section{Interpretation of higher order zero dynamics in the instantaneous rest frame}
\label{appB}
We show here that, at least in $1+1$, requiring $\vec B=0$ and $\vec H=0$ is equivalent to having constant values for the spatial part of the corresponding quantities of one order less in the instantaneous rest frame of the particle, \textit{i.e.} $\vec B=0$ in the lab frame is equivalent to constant acceleration in the rest frame, and so on. {   We do not have at present an equivalent result in $d+1$.}

In $(1+1)$ dimensions equations (\ref{cova}), (\ref{covb}), (\ref{covh}) boil down to   
\begin{eqnarray}
a^\mu &=& \left(\frac{\gamma^4}{c}{v}{A} , \gamma^4  {A}\right),\label{cova1}\\
j^\mu &=& \left(\frac{\gamma^5}{c}{v}{B}, \gamma^5  {B}\right),\label{covb1}\\
s^\mu &=& \left( \frac{\gamma^6}{c}{v}{H}, \gamma^6  {H}\right),\label{covh1}
\end{eqnarray}
with 
\begin{eqnarray}
A &=&  a,\label{ad1}\\
B &=&  b + 3 \frac{\gamma^2}{c^2} v a^2,\label{bd1}\\
H &=&  h + 10 \frac{\gamma^2}{c^2} vab + 3 \frac{\gamma^2}{c^2} a^3  + 18 \frac{\gamma^4}{c^4} v^2 a^3.
\end{eqnarray}

A general Lorentz transformation in $1+1$ is given by
\begin{equation}
\Lambda(v) = \begin{pmatrix}  
\gamma & -\gamma \frac{v}{c} \\ -\gamma \frac{v}{c} & \gamma
\end{pmatrix},\label{genL}
\end{equation}
and $a^\mu$, $j^\mu$, $s^\mu$ in the rest frame are then
\begin{eqnarray}
a^\mu_{\text{co}}&=&\Lambda(v) a^\mu=\begin{pmatrix} 0 \\ \gamma^3 a \end{pmatrix}
\equiv \begin{pmatrix} 0 \\  a_\text{co} \end{pmatrix},\label{a_co}\\
j^\mu_{\text{co}}&=&\Lambda(v) j^\mu=\begin{pmatrix} 0 \\ \gamma^4 b + 3 \frac{\gamma^6}{c^2} v a^2 \end{pmatrix}
\equiv\begin{pmatrix} 0 \\ b_\text{co} \end{pmatrix},\label{b_co}\\
s^\mu_{\text{co}} &=&  \Lambda(v) s^\mu
= \begin{pmatrix} 0 \\ \gamma^5 h + 10 \frac{\gamma^7}{c^2} v a b + 3 \frac{\gamma^7}{c^2} a^3 + 18 \frac{\gamma^9}{c^4}v^2a^3 \end{pmatrix}
\nonumber\\
&\equiv& \begin{pmatrix} 0 \\ h_\text{co} \end{pmatrix}.\label{h_co}
\end{eqnarray}
where $a_\text{co}$, $b_\text{co}$, $h_\text{co}$ represent the spatial part of the $1+1$ acceleration, jerk and snap, respectively, in the instantaneous rest frame.

A simple calculation shows that
\begin{align}
\frac{\dif}{\dif t}(a_\text{co}) &=\frac{\dif}{\dif t}(\gamma^3 a) = \gamma^3  b  +3 \gamma^2 \frac{\dif\gamma}{\dif t}  a\nonumber\\  & =
\gamma^3  b + 3 \gamma^5 \frac{ v  a}{c^2} a
=\gamma^3  B,
\end{align}
from which it follows that, in $1+1$, zero lab jerk  ($B=0$) implies that the spatial acceleration measured  in the rest frame is constant. Analogously, 
\begin{equation}\label{zero-snap}
\frac{\dif}{\dif t}(\gamma^4 b + 3 \frac{\gamma^6}{c^2} v a^2)
=\gamma^4  H,
\end{equation}
showing that, in $1+1$, zero lab snap ($H=0$) is equivalent to constant rest jerk.

The 1+1 Lorentz transformation (\ref{genL}) can be parameterized as
\begin{equation}
\Lambda(v)= \begin{pmatrix} \cosh\phi & -\sinh\phi \\ -\sinh\phi & \cosh\phi \end{pmatrix},
\label{PJ1}
\end{equation}
with $\phi(\tau)$ arbitrary. The velocity in the lab frame can be obtained as
\begin{equation}
v = c \tanh\phi,
\label{PJ2}
\end{equation}
and the derivative of the lab time with respect to the proper time is then
\begin{equation}
\frac{\dif t}{\dif \tau} = \gamma = \cosh\phi.
\label{PJ3}
\end{equation}
Deriving $v$ with respect to $t$ and using (\ref{PJ3}) to express the derivatives in terms of those of $\phi(\tau)$ with respect to $\tau$ (denoted by primes), one gets
\begin{eqnarray}
a &=& \frac{\dif v}{\dif t}= \frac{c}{\cosh^3\phi} \phi',\label{PJ4a}\\
b &=&\frac{\dif a}{\dif t} = -\frac{3c \sinh\phi}{\cosh^5\phi}  (\phi')^2 + \frac{c}{\cosh^4\phi}\phi'',
\label{PJ4b}\\
h &=& \frac{\dif b}{\dif t} = -\frac{3c}{\cosh^5\phi}(\phi')^3 + \frac{15c\sinh^2\phi}{\cosh^7\phi} (\phi')^3\nonumber\\  
&-&\frac{10c \sinh\phi}{\cosh^6\phi}\phi'\phi'' + \frac{c}{\cosh^5\phi}\phi''',
\end{eqnarray}
and so on.  Then,
\begin{enumerate}
	\item if $\phi(\tau)=\frac{v_0}{c}$, then $a=b=h=\ldots = 0$.
	\item if $\phi(\tau) = \frac{1}{c}(v_0 + a_0 \tau)$, then 
	\begin{eqnarray}
	a &=&\frac{1}{\gamma^3} a_0,\\
	b &=& -\frac{3v}{\gamma^4}\frac{a_0^2}{c^2} \neq 0,\\
	h &=& \frac{3(4\gamma^2-5)}{c^2\gamma^7}a_0^3 \neq 0,
	\end{eqnarray}
	but it follows from these equations that, in the rest frame,
	\begin{eqnarray}
	a_{\text{co}} &=& \gamma^3 a = a_0,\\
	b_{\text{co}} &=& \gamma^4 b + 3 \frac{\gamma^6}{c^2} v a^2 =0,\\
	h_{\text{co}} &=& \gamma^5 h + 10 \frac{\gamma^7}{c^2} v a b\nonumber\\ &+& 3 \frac{\gamma^7}{c^2} a^3 + 18 \frac{\gamma^9}{c^4}v^2a^3=0.
	\end{eqnarray}
	\item if $\phi(\tau) = \frac{1}{c}(v_0 + a_0 \tau+\frac{1}{2}j_0 \tau^2)$ one gets, in the rest frame of the particle,
	\begin{eqnarray}
	a_{\text{co}} &=& \gamma^3 a =  a_0+j_0 \tau,\\
	b_{\text{co}} &=& \gamma^4 b + 3 \frac{\gamma^6}{c^2} v a^2 = j_0, \\
	h_{\text{co}}  &=& \gamma^5 h + 10 \frac{\gamma^7}{c^2} v a b\nonumber \\ &+& 3 \frac{\gamma^7}{c^2} a^3 + 18 \frac{\gamma^9}{c^4}v^2a^3=0.
	\end{eqnarray}
	\item  if $\phi(\tau) = \frac{1}{c}(v_0 + a_0 \tau+\frac{1}{2}j_0 \tau^2+\frac{1}{6}s_0\tau^3)$,
	\begin{eqnarray}
	a_{\text{co}} &=&\gamma^3 a =  a_0+j_0 \tau+\frac{1}{2}s_0\tau^2,\\
	b_{\text{co}} &=& \gamma^4 b + 3 \frac{\gamma^6}{c^2} v a^2 = j_0+s_0\tau, \\
	h_{\text{co}}  &=& \gamma^5 h + 10 \frac{\gamma^7}{c^2} v a b\nonumber\\ &+& 3 \frac{\gamma^7}{c^2} a^3 + 18 \frac{\gamma^9}{c^4}v^2a^3= s_0.
	\end{eqnarray}
\end{enumerate}
and so on.

Notice that $v=c\tanh\phi(\tau)$ cannot be explicitly integrated with respect to $t$ to obtain $x(t)$.  An implicit solution is given in \cite{Russo:2009yd}
(see also \cite{Pons:2018lnw}) as
\begin{eqnarray}
x(\tau) &=& x_0 + c \int_0^\tau  \sinh\phi(s)\dif s,\\
t(\tau) &=& \int_0^\tau  \cosh\phi(s)\dif s.
\end{eqnarray}
The integrals on the right-hand sides can only be computed in terms of elementary functions, and inverted, for $\phi$ up to degree $1$ in $\tau$.

\vfill

\end{appendix}

\bibliography{LyeSym}

\end{document}